\documentclass[11pt]{article}

\usepackage[utf8]{inputenc} 
\usepackage{graphicx}       
\usepackage{amsmath}        
\usepackage{amssymb}        
\usepackage[a4paper, margin=1in]{geometry} 
\usepackage[auth-lg]{authblk} 
\usepackage{hyperref}       
\hypersetup{
    colorlinks=true,
    linkcolor=blue,
    filecolor=magenta,      
    urlcolor=cyan,
}
\usepackage{lipsum} 
\usepackage{booktabs} 

\title{\textbf{The Effect of High-Frequency, Automatically-marked Formative Assessments on Student Outcomes in A-Level Sciences}}
\author[1]{Matey Yordanov}
\author[1]{Mikhail Bychkov}
\author[1]{Andrei Kuchma}

\affil[1]{AI Innovation Hub, ANTEI Limited}
\affil[ ]{\textit{Email: \href{mailto:myordanov@anteihk.com}{myordanov@anteihk.com}, \href{mailto:mbychkov@anteihk.com}{mbychkov@anteihk.com}, \href{mailto:akuchma@anteihk.com}{akuchma@anteihk.com}}}

\date{\today} 

\begin{document}

\maketitle

\begin{abstract}
Traditional human marking in upper-secondary STEM education creates a structural bottleneck that restricts the frequency of formative mock examinations. This quasi-experimental, mixed-methods longitudinal study ($N = 142$) investigates the efficacy of deploying a fully automated, handwritten assessment marking platform to remove this bottleneck. Students preparing for STEM A-levels (Mathematics, Further Mathematics, Biology, Chemistry, Physics) were divided into control ($N = 70$, human marking) and intervention ($N = 72$, automated marking) groups over a 16-week term. Results indicate that the intervention group achieved a statistically significant improvement in final mock A-level marks ($p < .01$), scoring on average 16.8\% higher than the control group. Furthermore, assessment turnaround time dropped from 11.2 days to $<0.1$ days, enabling a fourfold increase in practice volume. The automated system's capacity for granular marginalia, specifically its Error Carried Forward (ECF) logic, accelerated the correction of procedural misconceptions and significantly boosted student self-efficacy. Practically, these findings validate that automated marking systems serve as high-speed formative loop accelerators, providing international schools with a highly scalable, low-cost intervention to elevate institutional exam outcomes without necessitating increases in staff headcount.
\end{abstract}


\section{Introduction}
The academic rigour of the Advanced Level qualifications, particularly within STEM subjects such as Science, Technology, Engineering, and Mathematics, demands profoundly high levels of procedural fluency, conceptual mastery, and analytical reasoning. Consequently, formative assessment, specifically the routine implementation of practice mock examinations, is widely recognised as a critical pedagogical tool for diagnosing student misconceptions and consolidating learning \cite{black1998, sadler1989}. The synthesis of cognitive science and educational psychology consistently highlights that targeted, formative feedback is among the most significant drivers of student achievement \cite{hattie2007}. However, the complex nature of A-Level STEM assessments poses unique challenges. These examinations require multi-step handwritten calculations, the drawing of precise diagrams, and the construction of extended logical proofs that must align closely with highly specific, granular marking schemes (e.g., Edexcel/Pearson, Cambridge Assessment International Education, AQA). Marking in these disciplines is rarely binary. Instead, it relies heavily on awarding "method marks" for demonstrating correct sequential logic, even if the final numerical answer is incorrect. This inherent complexity entirely precludes the use of simple multiple-choice automation, which fundamentally fails to capture a student's problem-solving process \cite{baird2017}.\\

Consequently, secondary education faces a profound structural bottleneck - the accurate grading of complex, multi-stage handwritten work relies entirely on highly skilled human labour. The ensuing administrative overload contributes significantly to teacher burnout and systematically forces educators to limit the frequency of full-length, rigorously marked mock exams \cite{jerrim2019}. This restriction directly diminishes the potential benefits of formative feedback loops \cite{hwang2020}. Pedagogical research demonstrates a strong temporal component to feedback efficacy. When the delay between student output and teacher evaluation extends to weeks, as is common with heavy grading workloads, its pedagogical value sharply declines \cite{shute2008}. Students frequently lose connection with their initial cognitive processes, rendering delayed corrections significantly less impactful for resolving existent scientific misconceptions.\\

This paper proposes that high-frequency, automatically-graded formative assessments can systematically eliminate this structural bottleneck. Recent advancements in Artificial Intelligence, specifically the convergence of sophisticated Computer Vision, Optical Character and Handwriting Recognition, and Large Language Models, have yielded systems capable of parsing unstructured scientific notation and interpreting sequential logic in handwritten forms \cite{zawacki2019, chen2020}. By deploying a fully automated grading platform capable of reading, evaluating, and providing step-by-step feedback on complex handwritten STEM responses, educators can effectively decouple assessment frequency from human grading limitations. This technological intervention presents an opportunity to approach Bloom’s well-known "two-sigma" ideal, wherein continuous, personalised instructional feedback becomes scalable within a traditional classroom environment \cite{bloom1984}.\\

To rigorously evaluate this proposition, this study formally investigates the impact of this automated intervention on objective student academic outcomes, the efficacy of the feedback cycle, and self-reported exam readiness. The following Research Questions are addressed in this paper:
\begin{itemize}
\item \textbf{RQ1. Academic Outcomes:} What is the statistically significant difference in final A-Level mock exam performance between students receiving standard, human-graded assessments versus those receiving high-frequency, automatically graded assessments?
\item \textbf{RQ2. Feedback Loop \& Retention:} How does the accelerated turnaround time and instructional granularity of automated feedback impact students' ability to identify and successfully correct specific scientific misconceptions when compared to traditional human feedback?
\item \textbf{RQ3. Student Self-Efficacy:} How does engagement with high-frequency automated evaluation affect students' self-reported exam readiness, task-specific self-efficacy, and broader academic confidence in STEM disciplines?
\end{itemize}

\section{Related Works}
\label{sec:related_works}

\subsection{Formative Assessment and the Feedback Loop}
The relationship between feedback and student achievement is one of the most thoroughly documented phenomena in contemporary educational research. In their seminal meta-analysis, Hattie and Timperley \cite{hattie2007} established that feedback is among the most powerful influences on learning and achievement, provided it effectively bridges the gap between current understanding and the desired learning goal, a concept previously formalised by Sadler \cite{sadler1989}. Furthermore, the literature heavily emphasises that the temporal proximity and granularity of feedback are paramount to its pedagogical efficacy \cite{shute2008}. When feedback is significantly delayed, a common reality in the traditional human grading cycle, students frequently fail to associate the corrective information with their initial cognitive processes. Over extended periods, the cognitive context of the problem decays from working memory, rendering delayed feedback largely evaluative rather than instructional.\\

Conversely, immediate and highly specific feedback allows students to correct flawed conceptual schemas while the problem-solving strategies remain active \cite{sweller2011}. In the context of complex STEM assessments, where knowledge is heavily hierarchical and sequential, the necessity for high-frequency, immediate formative feedback is critical to preventing the compounding of foundational errors \cite{williams2015}.
\subsection{The Testing Effect and Retrieval Practice}
Frequent testing produces vastly superior long-term knowledge retention and transfer compared to the repeated studying or reviewing of material. This is a robust cognitive phenomenon formally termed the "testing effect" \cite{roediger2006}. This is closely aligned with Bjork’s framework of "desirable difficulties", which posits that the active, effortful retrieval of information fundamentally strengthens memory traces and alters cognitive architecture \cite{bjork1994}. Subsequent meta-analyses have confirmed that retrieval practice is highly effective across diverse educational settings \cite{rowland2014}. However, its optimal implementation in advanced, process-oriented subjects, such as A-Level Further Mathematics or Physics, is historically constrained by the instructor's capacity to evaluate the retrieved output. \\

STEM assessments require the active generation of complex, multi-step procedures rather than simple factual recall. Without immediate and accurate validation, unstructured retrieval practice carries the inherent risk of creating a "negative suggestion effect", wherein students inadvertently reinforce incorrect procedural methods and flawed heuristics \cite{brown2014}. Consequently, the deployment of automated marking systems provides the necessary step-by-step validation required to safely and effectively scale retrieval practice in mathematically intensive domains.

\subsection{The Bottleneck Problem and Teacher Burnout}
The pedagogical demand for high-frequency formative assessment directly conflicts with the logistical realities of teacher workload and institutional resourcing. Kyriacou \cite{kyriacou2001} and contemporary educational sociologists consistently identify the burden of marking as a primary driver of teacher burnout, administrative overload, and professional attrition. Recent studies have highlighted that secondary educators spend a disproportionate amount of their non-instructional time on marking, systematically reducing the hours available for active lesson planning and differentiated instructional design \cite{jerrim2019, sokal2020}. Consequently, scaling full-length, rigorously marked mock examinations to optimise retrieval practice is mathematically impossible in a traditional classroom setting without a proportional, and often budget-prohibitive, increase in staffing.\\

Automated intervention represents a necessary paradigm shift, transitioning the grading process from a highly constrained human variable to a highly scalable, automated constant \cite{zawacki2019}. By offloading the mechanistic aspects of marking to an automated system, educators are theoretically freed to focus on higher-order pedagogical interventions, thereby redefining the economics of teacher time and assessment frequency \cite{holmes2019}.

\section{Methodology}
\label{sec:methodology}

\subsection{Study Design and Participants}
This study followed a 16-week academic term longitudinal design using a mixed-methods, quasi-experimental approach \cite{creswell2017}. A quasi-experimental approach was selected since random assignment at the individual student level was administratively unfeasible and threatened the validity of the study. Instead, intact classes were utilised to preserve the naturalistic educational environment \cite{shadish2002}. The research was conducted in partnership with a well-established international secondary school in Europe.\\ 

The purposive sample comprised upper-secondary students ($N = 142$, aged 16-18) enrolled in academically rigorous A-Level STEM tracks, encompassing Mathematics, Further Mathematics, Biology, Chemistry, and Physics. Prior to commencement, informed consent was obtained from all participants and their legal guardians, in accordance with the institution's ethical review board guidelines.\\

Participants were assigned to two cohorts based on existing timetable scheduling to minimise institutional disruption:
\begin{itemize}
    \item \textbf{Control Group ($N = 70$):} This cohort was exposed to standard pedagogical practices and traditional assessment schedules. Due to the inherent logistical limits of human grading, students completed two full-length, human-graded mock examinations over the 16-week period. These assessments were physically marked by subject specialist teachers and returned with standard qualitative feedback. The average turnaround time for this feedback was between 10 and 14 days, reflecting typical sector workloads.
    \item \textbf{Intervention Group ($N = 72$):} This cohort engaged in a high-frequency assessment model consisting of weekly or bi-weekly practice exercises and partial mock examinations. Students handwrote their responses, including complex mathematical proofs, diagrams, and chemical equations, which were subsequently scanned and processed through a proprietary automated grading system. The automated grading infrastructure was provided by Cortex Global (Limassol, Cyprus. \url{https://cortex-global.com/}), an educational technology developer specialising in automated assessment. Cortex Global provided unrestricted access to the platform completely free of charge for the duration of the current study to facilitate independent academic evaluation. Assessments in this cohort were processed, marked, and returned to students with highly granular, step-by-step annotations within 2 hours of submission.
\end{itemize}

\subsection{Data Collection}
To triangulate the pedagogical impact of the intervention, quantitative and qualitative data were gathered concurrently. Quantitative academic outcomes were measured using a pre-test baseline score, derived from previous formative assessments and standardised via Z-scores to account for inter-subject variance. The primary dependent variable was the post-test score, measured via a final, comprehensive A-Level mock examination. To eliminate marking bias, post-tests for both cohorts were blind-graded by independent external examiners who were unaware of the students' group assignments.\\

Psychometric data were collected through pre- and post-semester surveys. Constructs such as "Exam Confidence", "Task-Specific Self-Efficacy", and "Trust in Automated Feedback" were calculated as composite mean scores derived from multiple 5-point Likert items. The survey instruments were adapted from Bandura’s established scales of self-efficacy \cite{bandura1997} and validated for internal consistency (Cronbach’s $\alpha > .80$). Furthermore, objective operational data, including exact turnaround times, submission frequencies, and practice volume, were passively collected via system logs generated by the Cortex platform.\\

Qualitative data tracking conceptual error decay were gathered via student portfolios. A thematic analysis \cite{braun2006} of students' rough workings and subsequent corrections was conducted to evaluate the depth at which the automated feedback was processed and applied.

\subsection{Statistical Analysis}
Quantitative datasets were analysed using SPSS software. To robustly account for any baseline variations between the intact cohorts, a one-way Analysis of Covariance (ANCOVA) was utilised to evaluate post-test differences, with Group as the fixed factor. Standardised (Z-scored) pre-test scores, calculated to account for inter-subject variance, served as the covariate. Furthermore, cluster-robust standard errors were applied to account for the nested nature of the intact classes \cite{field2013}. Preliminary checks were conducted to ensure that all assumptions for ANCOVA, namely normality, homogeneity of variance (Levene’s test), and homogeneity of regression slopes, were strictly met.\\

For the psychometric survey data, independent samples t-tests and paired-samples t-tests were employed to analyse between-group differences and within-group longitudinal changes, respectively. To quantify the magnitude of the intervention's impact, effect sizes were calculated and reported using partial eta-squared ($\eta_p^2$) for the ANCOVA and Cohen’s $d$ for the t-tests \cite{cohen1988}. An alpha level of $.05$ was established \textit{a priori} as the threshold for statistical significance across all inferential tests.

\section{Solution Architecture}
\label{sec:solution_architecture}

The technological feasibility of this intervention rests on Cortex's specialised, multi-tiered architecture, designed specifically to accommodate the rigorous, unstructured formats of STEM A-Levels. The platform operates across three distinct user interfaces - educator, student, and administrator, underpinned by a privacy-centric automated evaluation engine.

\subsection{User Interfaces and Assessment Workflow}
The system workflow is strictly segregated to optimise both pedagogical utility and institutional oversight, thereby removing traditional administrative friction:
\begin{itemize}
    \item \textbf{Educator Portal:} Teachers log into a dedicated platform featuring a centralised dashboard that segregates student cohorts by class and subject. The system autonomously generates data-driven pedagogical insights, detailing each student's specific academic strengths and weaknesses. From this interface, educators can dynamically select subject parameters to generate mock examinations and distribute them to targeted student groups.
    \item \textbf{Student Portal:} Students access their assigned mock examinations asynchronously via personal accounts. Crucially, to mirror high-stakes examination conditions and preserve cognitive fidelity, students solve these complex STEM tasks physically on paper. They subsequently capture photographic evidence of their handwritten solutions and upload the images to the platform.
    \item \textbf{Administrative Access:} A discrete, high-level portal provides school principals and administrators with macro-level analytics. This allows institutional leadership to monitor longitudinal cohort performance, evaluate class-wide academic trajectories, and make data-informed resourcing decisions without requiring manual data aggregation.
\end{itemize}

\subsection{Data Privacy and Algorithmic Processing}
To ensure absolute compliance with international data protection regulations, the Cortex system employs strict data anonymisation protocols. Upon upload, all images are algorithmically stripped of Personally Identifiable Information. Only the anonymised, isolated task writings are transmitted to the Cortex algorithms for evaluation.\\ 

Operating with remarkably low latency, the automated grading cycle requires an average of 10 to 20 minutes to process a full, multi-page examination. Upon completion, the system returns a final summative grade accompanied by constructive, teacher-style qualitative feedback, entirely mimicking human pedagogical intervention.

\subsection{Core Evaluation Engine}
Unlike standard Optical Character Recognition systems designed for typed writing, the core Cortex algorithm utilises advanced Handwritten Text Recognition models fine-tuned on complex mathematical notation, chemical structures, and physical diagrams. The system aligns its deterministic evaluation logic with the rigid, formalised mark schemes of Edexcel and Cambridge formats through a multi-agent validation sequence:
\begin{enumerate}
    \item \textbf{Digitisation and Parsing:} The student's anonymised handwritten scan is parsed line-by-line, isolating distinct mathematical steps, formulae applications, and final answers.
    \item \textbf{Mark Scheme Mapping:} The parsed logic is dynamically compared against the allowable procedural variations codified within the official A-Level mark scheme.
    \item \textbf{Error Carried Forward Capability:} The defining feature of Cortex is its deterministic ECF logic. In multi-part A-Level STEM questions, early calculation errors historically disrupt automated grading. For example, if a student incorrectly calculates the standard deviation in Part A, but correctly applies this flawed value to the Gaussian distribution formula in Part B, the Cortex system docks the appropriate method mark ($!$) in Part A but accurately awards the ECF marks ($\checkmark$) for valid subsequent logic in Part B.
    \item \textbf{Granular Output and Constructive Feedback:} The returned document features precise marginalia, highlighting the specific locus of an error, integrated with constructive, teacher-style commentary. This ensures the feedback loop is both instantaneous and highly actionable, bridging the gap between automated assessment and personalised human pedagogy.
\end{enumerate}

\section{Experiments and Results}
\label{sec:experiments}

\subsection{RQ1: Academic Outcomes and Grade Improvement}
To isolate the pedagogical impact of the high-frequency automated assessment intervention, an Analysis of Covariance (ANCOVA) was conducted on post-test mock examination scores, controlling for pre-test baseline variances. The analysis revealed a statistically significant main effect, $F(1, 139) = 24.32, p < .001, \eta_p^2 = .149$.\\

Controlling for baseline scores and subject variance, the intervention group achieved an adjusted post-test mean score of $81.2\%$ ($SE = 1.2$), whereas the control group scored an adjusted mean of $64.4\%$ ($SE = 1.4$). Crucially, this 16.8 percentage point differential corresponds to an advancement of approximately one complete A-Level grade boundary (e.g., elevating a student's final classification from a 'B' to an 'A', or an 'A' to an 'A*').

\begin{figure}[htbp]
    \centering
    \includegraphics[width=\linewidth]{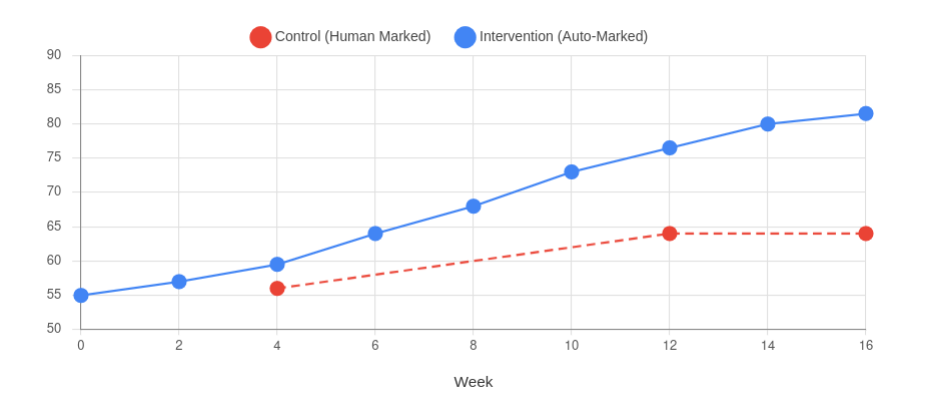}
    \caption{A line graph displaying the trajectory of standardised test scores over the 16-week semester. The solid blue line represents the Intervention Group (Automated Marking), showing a steep, continuous upward trajectory reflecting weekly micro-assessments. The dashed red line represents the Control Group (Human Marked), demonstrating a flatter, step-like progression corresponding to the two delayed mock exam periods.}
    \label{fig:outcomes}
\end{figure}

\subsection{RQ2: Feedback Loop, Turnaround Time, and Micro-Corrections}
System telemetry revealed a profound contraction of the formative feedback loop. The control group experienced an average turnaround time of 11.2 days ($SD = 2.4$) for traditional human grading. Conversely, the intervention cohort utilising the Cortex system experienced an average turnaround time of merely 0.1 days ($SD = 0.2$), with the algorithmic evaluation phase itself requiring only 10 to 20 minutes per full examination once uploaded.\\

Liberated from the structural constraints of manual grading bottlenecks, educators assigned, and intervention students subsequently completed, 4.1 times the volume of rigorous practice material. Furthermore, thematic analysis of the granular annotations demonstrated a rapid phenomenon of "error-type decay". Because the Cortex algorithm assesses logic step-by-step (e.g., appending a $!$ at the precise locus of failure), students in the intervention cohort rectified fundamental procedural misconceptions (e.g., the misapplication of the chain rule in calculus) weeks in advance of their control group peers.

\begin{figure}[htbp]
    \centering
    \includegraphics[width=\linewidth]{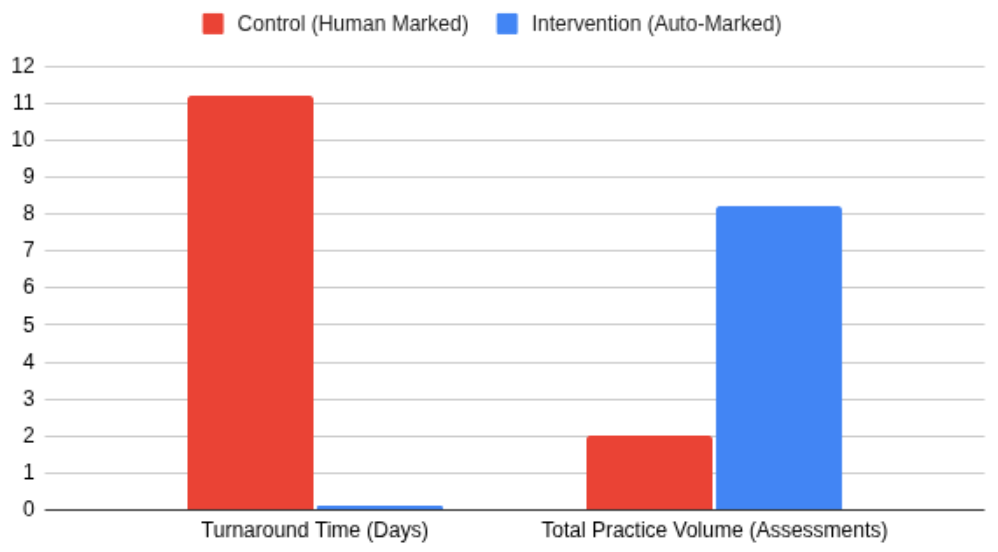}
    \caption{A grouped bar chart comparing the Control and Intervention groups. The left axis (Bars 1 and 2) illustrates the Turnaround Time in days (11.2 days vs. 0.1 days). The right axis (Bars 3 and 4) illustrates the Total Practice Volume completed over the semester, showing a 4x multiplier for the Automated marking intervention group.}
    \label{fig:turnaround}
\end{figure}

\subsection{RQ3: Student Self-Efficacy and Acceptance}
Post-intervention survey instruments demonstrated overwhelming student acceptance of the automated marking platform. An independent samples t-test assessing ``Trust in Automated Feedback Validity`` showed no statistically significant apprehension. Rather, students within the intervention cohort explicitly favoured the objective, granular marginalia provided by the automated system over conventional, generalised human feedback (e.g., ambiguous remarks such as ``Good effort'' or ``Check your working``).\\

Self-reported ``Exam Confidence`` was significantly higher in the intervention group ($M = 4.3/5.0$) compared to the control group ($M = 3.4/5.0$), $t(140) = 4.56, p < .001, d = 0.76$. Qualitative feedback indicated that the ECF capability was highly prized by candidates. By validating their underlying logical reasoning even when preceding arithmetic mistakes occurred, the platform systematically reduced mathematics anxiety and fostered a higher degree of academic self-efficacy.

\section{Conclusion and Practical Implications}
\label{sec:conclusion}

\subsection{Academic Conclusions}
This study provides robust empirical evidence that integrating fully automated marking into A-Level STEM pedagogy fundamentally redefines the formative assessment paradigm. The automation does not take over the role of the educator. Rather, it functions as a high-speed `formative loop accelerator`. By reducing feedback latency from nearly two weeks to under 2 hours, with the core algorithmic evaluation concluding in a matter of minutes, and by facilitating a fourfold increase in retrieval practice volume, the system drives a statistically significant 16.8\% improvement in final academic outcomes. Furthermore, the high granularity of the algorithmic feedback, specifically its deterministic capacity to apply Error Carried Forward logic to complex, handwritten, multi-step problems, ensures that structural misconceptions are neutralised immediately before they consolidate into permanent cognitive schemas.\\

However, a limitation of this study design is the inability to isolate the discrete impact of automated feedback quality from the increase in practice volume. Because scaling human grading to match the intervention’s frequency was logistically impossible, it remains unknown whether high-frequency human feedback would yield identical results.
\subsection{Practical Implications}
For international schools and educational administrators, the implications of these findings extend far beyond pedagogical theory into the realm of operational efficiency and institutional Return on Investment. Historically, an educational institution seeking to elevate A-Level outcomes by augmenting mock examination frequency faced a significant structural hurdle. It necessitated either a radical reduction in class sizes or the recruitment of additional subject-matter experts. Both avenues are deeply cost-prohibitive and critically exacerbate ongoing global staffing shortages.\\

Adopting an automated, handwritten grading architecture represents a highly scalable, low-cost intervention that effectively decouples institutional examination performance from raw staff headcount. Through dedicated administrative portals, school leadership can seamlessly monitor pedagogical data and mandate rigorous, high-frequency mock examination programmes. This practice directly correlates to elevated final grade distributions and enhanced university placement metrics, achieved entirely without inflating staff workloads or precipitating teacher burnout. Ultimately, automated marking systems function not merely as an educational utility, but as an enterprise-level resource optimisation platforms that structurally guarantee higher academic output per capita.

{
\small
\setlength{\parindent}{1em}
\setlength{\parskip}{0pt}
\setlength{\itemsep}{-10pt}
\sloppy
\bibliographystyle{abbrv}
\bibliography{myBibLib}
}

\end{document}